\documentclass[%
twocolumn,
nofootinbib,
amsmath,amssymb,
aps,
showkeys
]{revtex4-1}
\usepackage{graphicx}
\usepackage{dcolumn}
\usepackage{bm}
\usepackage{color}

\begin{document}

\preprint{APS/123-QED}

\title{Temporal inactivation enhances robustness in an evolving system}

\author{Fumiko Ogushi}
\email{ogushi.fumiko.54n@st.kyoto-u.ac.jp}
\affiliation{
Kyoto University Institute for Advanced Study, Kyoto University, Yoshida Ushinomiya-cho, Sakyo-ku, Kyoto, 606-8501, JAPAN
}%
\affiliation{
Center for Materials research by Information Integration, National Institute for Materials Science, 1-2-1 Sengen, Tsukuba, Ibaraki 305-0047, JAPAN
}%
\author{J\'{a}nos Kert\'{e}sz}
\email{KerteszJ@ceu.edu}
\affiliation{
Department of Network and Data Science, Central European University, 1051 Budapest, Hungary
}%
\affiliation{
Institute of Physics, Budapest University of Technology and Economics, 1111 Budapest, Hungary
}
\author{Kimmo Kaski}%
\email{kimmo.kaski@aalto.fi}
\affiliation{
Department of Computer Science, Aalto University School of Science, P.O. Box 15500, Espoo, Finland
}%
\affiliation{
The Alan Turing Institute, British Library, 96 Euston Road, London NW1 2DB, UK
}%
\author{Takashi Shimada}
\email{shimada@sys.t.u-tokyo.ac.jp}
\affiliation{%
Mathematics and Informatics Center, The University of Tokyo}
\affiliation{
Department of Systems Innovation, Graduate School of Engineering,
The University of Tokyo,
7-3-1 Hongo, Bunkyo-ku, Tokyo, 113-8656, JAPAN
}%

\date{\today}

\begin{abstract}
We study the robustness of an evolving system that is driven by successive inclusions of new elements or constituents with $m$ random interactions to older ones. Each constitutive element in the model stays either active or is temporarily inactivated depending upon the influence of the other active elements. If the time spent by an element in the inactivated state reaches $T_W$, it gets extinct.
The phase diagram of this dynamic model as a function of $m$ and $T_W$ is investigated by numerical and analytical methods and as a result both growing (robust) as well as non-growing (volatile) phases are identified. It is also found that larger time limit $T_W$ enhances the system's robustness against the inclusion of new elements, 
mainly due to the system's increased ability to reject ``falling-together'' type attacks. 
Our results suggest that the ability of an element to survive in an unfavorable situation for a while, either as a minority or in a dormant state, could improve the robustness of the entire system.
\end{abstract}

\pacs{Valid PACS appear here}
\maketitle

\section{\label{sec:level1}Introduction}
The robustness of a system with many interacting elements or constituents under successive addition of new elements is an essential question for 
understanding the behaviour of various complex real world systems, that are often called ecosystems~\footnote[1]{Here the term ``ecosystem'' is used in a rather general sense to mean biological ecosystems but also diverse economical and social systems of individuals and institutions.}. In these systems the interactions between elements can be competitive or co-operative in nature such that the fitness of its elements or species can be strengthened or weakened by them, possibly causing the species getting extinct. This problem calls for a network theoretic approach, where the constituents of the system are the nodes of a dynamical network and the interactions are the links between them. Then the rephrased question is about the evolution of such a network of nodes under the condition that new nodes with different kinds of links are introduced. 
If the network can grow, then the evolving system it describes is considered robust, otherwise the system does not grow and is considered volatile. 
This way, we believe that the network approach can be used and be versatile in investigating various aspects of robustness for wide range of different systems.

Earlier it has been shown that in a simple model setting, where directed random positive and negative interactions characterize the system and the fitnesses of nodes (i.e. species) are identified with their strengths, when the links per node ratio 
---serving as a critical parameter--- 
remains within a certain range, the system is robust~\cite{Shimada2014SREP}. 
This mechanism and the resulting phase diagram of the growth of the system were found to be universal, i.e., this feature is shared among a variety of models like the one with different distributions of interaction weights and with constant or random number of links introduced with the new nodes~\cite{Shimada2015MABS} and even with different bidirectional correlations~\cite{Ogushi2017SREP}. While the range of robustness may be influenced by the details of the model, e.g., the mutuality in the interactions increasing it, the overall picture remains the same. 

An alternative way to study the problem of robustness in complex interacting systems is population dynamics based approach as often done in theoretical ecology~\cite{book_ETG_MaynardSmith, book_MathematicalBiology_Murray, book_Ecology_Harper}. 
Such a framework enables more complex dynamics and is flexible with respect to allowing different states of the species, but unlike in the network approach the inclusion of topological constraints are less straightforward in the population dynamics approach. 
Our aim here is to contribute to the convergence of these different approaches by including complex temporal features of interactions into the network models. 
\begin{figure}[b]
\centering 
\includegraphics[width=0.5\textwidth]{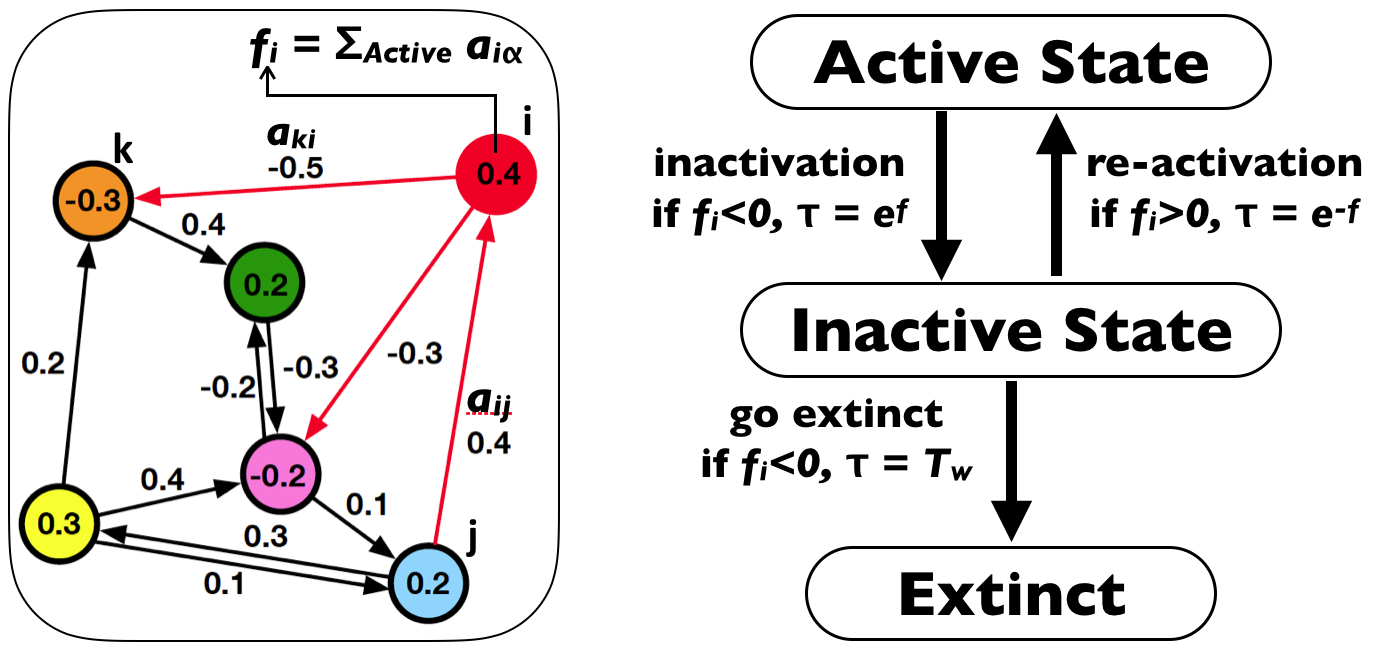}
\caption{
Introduction of the inactive state (dormancy) before the extinction, to our graph-dynamics framework. Less fit species is inactivated faster, and better fit species in inactive state is reactivated faster. The time limit of dormancy till extinction is, in contrast, uniformly set to $T_W$.
}
\label{fig_model_dormancy}
\end{figure} 

In population dynamics models, less fit species become minor in their population which in general makes that species almost irrelevant to the other species before that really gets extinct. 
For example, in the well adopted (generalized) Lotka-Volterra model~\cite{Taylor1988A, Taylor1988B} and replicator dynamics model~\cite{Tokita1999}, the trajectory starting from a feasible initial state (i.e. all population variables are positive~\cite{Roberts1974}) never touches $0$ within finite time. Therefore, 
a threshold is generally introduced to model extinction. This is a simplified treatment of the Allee effect~\cite{WhatIsAlleeEffect} about the weakening of the fitness in small populations, or rather direct modelling of the negative effect of demographic stochasticity~\cite{DemographicStochasty_Hastings2008, YohsukeRikvold_DemographicStochasticity_PRE2010}. In summary, these observations and the related approaches suggest that the population size of less fit species and its temporal derivative becomes very small before extinction and the process is often lengthy. 
Furthermore, the adaptive nature of foraging and other interactions
at the population level and at the individual level~\cite{book_Ecology_Harper, LR_Ciclid_Hori1993Science, AdaptiveForagingOfGrasshoppers_PNAS2000, Kondoh2003, Takeuchi_SREP2017,  RapidEvolutionLizard_2014} make such very minor species effectively even more invisible for other species. 
Therefore it seems plausible to include an ``inactive state'' into the set of possible states for handling such weakened populations. Species in such an inactive state, i.e. close to extinction, could be revived or reactivated within a frame of time if the circumstances would sufficiently improve.

The introduction of inactive state can be also regarded as modeling {\it dormancy}, which is broadly observed in biological ecosystems, such as in case of hibernation and surviving in seed, spore, or bacterial spore~\cite{hibernation_review_Andrews2007, plankton_dormancy_2012}. 
From the evolutionary point of view hibernation or dormancy is favorable as it enables survival under scarce conditions. Therefore, we expect that this new component if considered in the framework of network models will increase the robustness of the system, which in turn should be reflected in the increase of the growth region in the phase diagram.

The paper is organized such that in the next section we describe our network based model of evolutionary system of species capable of being temporarily inactive. This is followed with a comprehensive account and analysis of computational modeling results to map out the phase diagram of the evolutionary system. Then we draw conclusion and present discussions.

\section{\label{sec:level2}Model}
As we consider the ecosystems of being composed of connected species, we have devised our model being a network of nodes (or species) connected by unidirectional links with weights, as illustrated schematically in FIG. \ref{fig_model_dormancy}. Here the nodes represent species of animals of some sort and the links different types of directed influences between the pairs of species. The strength of the influence of species $j$ on species $i$ is denoted by the weight of the unidirectional link from node $j$ to node $i$, i.e. $a_{ij}$. These weights can be either positive or negative. Each species has its ``fitness'', which is simply given by the sum of its incoming interactions from other species in the system, i.e., $\displaystyle f_i = \sum_j^{incoming} a_{ij}$. A species can survive as long as its fitness is greater than zero. 
The species with non-positive  fitness, which in our previous model went instantaneously extinct, will in the present model be {\it inactivated} after its fitness-dependent waiting time $\tau = {\rm e}^{f}$, i.e. species in worse situation is inactivated faster. The inactivated species looses its influence on other species thus we will neglect the links out of those for the calculation of fitness. If the surrounding community of an inactivated species changes and the fitness of an inactivated species becomes positive, the species is reactivated (waking up from dormancy). The waiting time of this reactivation process is also assumed to be fitness-dependent: $\tau = {\rm e}^{-f}$. The slowest process among the microscopic dynamics is the inactivation and reactivation of solitary species ($f = 0$). The duration of these processes, $\tau = 1$, gives the unit of time to this otherwise timescale-less model. 
Although it is known that some species can maintain its dormancy for quite a long time~\cite{KumamushiSurvives30years}, the period has generally a limit. In the following, we introduce a uniform time-limit parameter $T_W$. A species that has spent $T_W$ of continuous time in the inactive state with non-positive fitness gets extinct. The extinct species and its incoming and outgoing links are removed permanently. 
Note that the present model with dormancy reduces to the original model at $T_W = 0$. 
A pseudo-code style description of the entire dynamics is available in the Appendix.
\begin{figure}[bth]
\centering 
\includegraphics[width=0.5\textwidth]{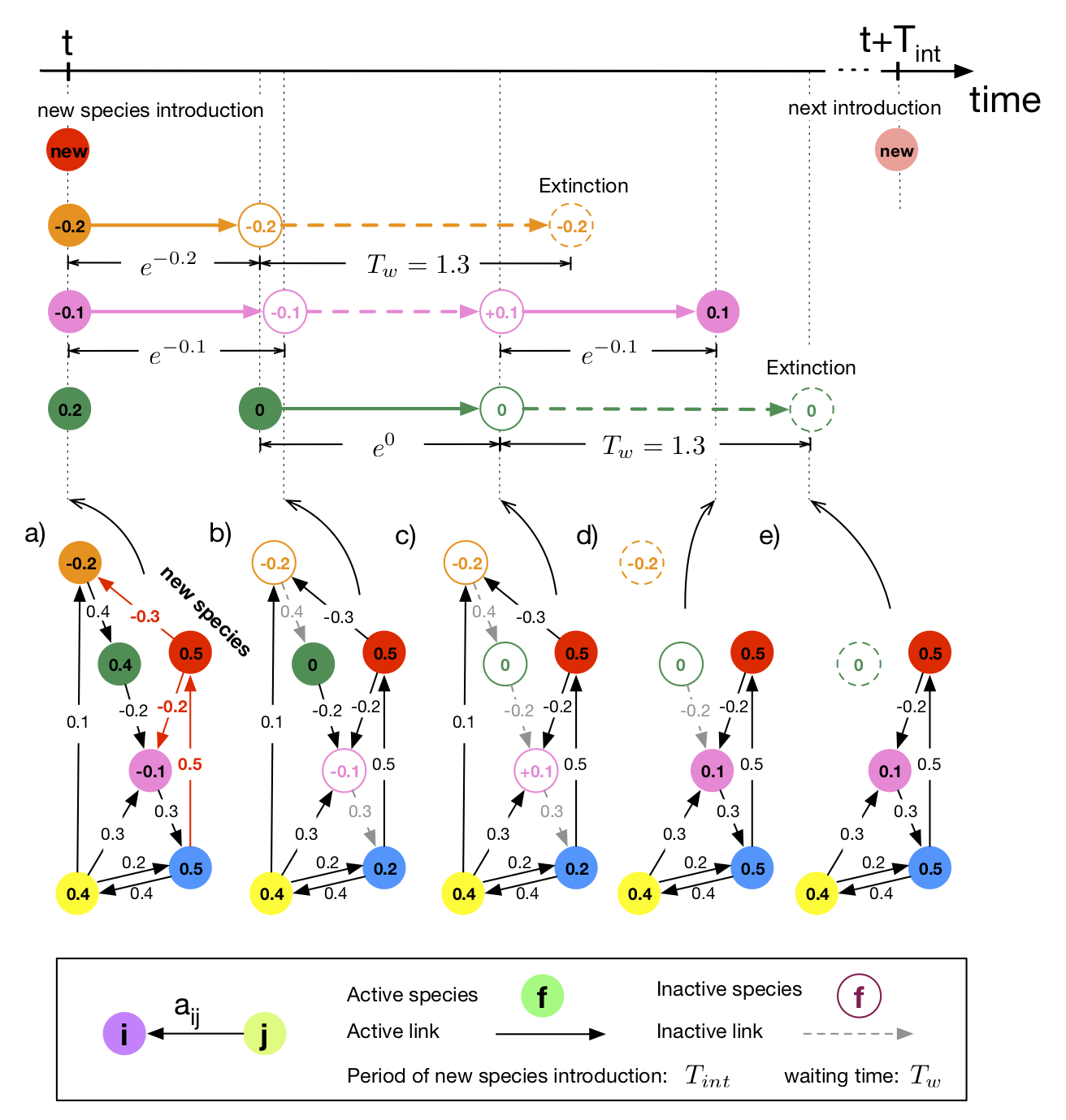}\caption{
A temporal evolution of the model with inactivation (dormancy) and reactivation (revival), after inclusion of new species. 
(a): Introduction of a new species (red), which makes the fitness of two species (orange and magenta) negative. 
Each of these two species will be inactivated after its fitness-dependent duration: $\tau = \exp\{f_i/f_0\}$. 
(b): Inactivation of the species with worse fitness (orange) takes place first and then the other species (magenta) is inactivated, which  makes the fitness of another species (green) non-positive. Inactivated species is given $T_W$ of waiting time till it will go extinct. 
(c): Green species is inactivated before any of other inactive species goes extinct. This change makes the fitness of the inactive species (magenta) positive. 
(d): Magenta species is reactivated after a fitness-dependent waiting time $\tau = \exp\{-f_i/f_0\}$. 
Meanwhile, the orange species have spent $T_W$ of time in the inactivated state and hence gone extinct: the orange species and the interactions from and to it are deleted. 
(e): Green species goes extinct. This does not change the sign of fitness of any species in the community. Therefore, after the extinction of green species, the system finally reaches to a new persistent state i.e. all the species are in the active state and have positive fitnesses.
Nothing will happen for a community in a persistent state, until the next new species is introduced at $t + T_{\rm int}$.
}
\label{fig_model_detail}
\end{figure}

\begin{figure}[bth]
\centering 
\includegraphics[width=0.4\textwidth]{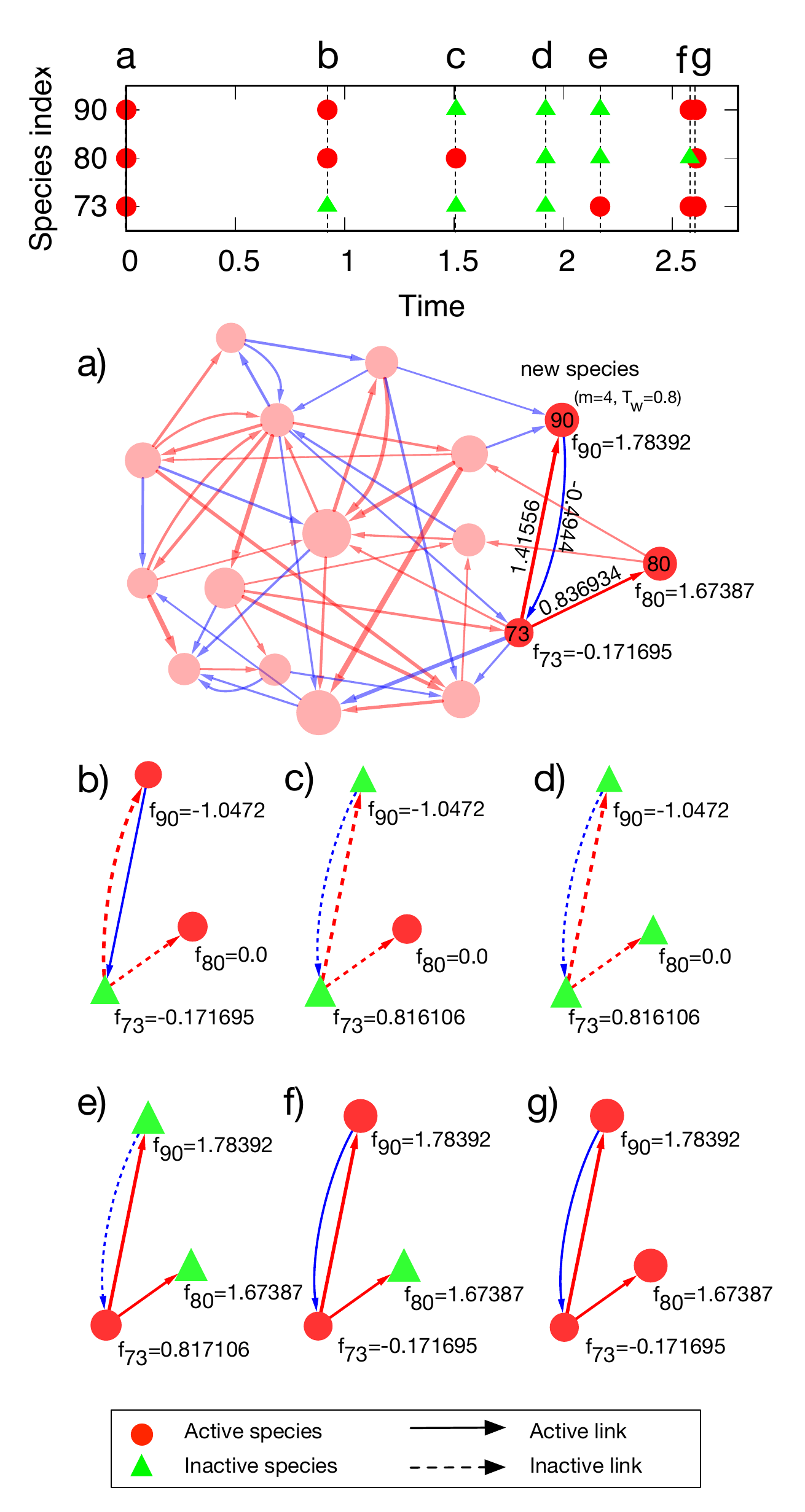}
\caption{
A limit cycle observed in an emergent system in the present model with inactivation and revival processes. 
}
\label{fig_blinking_behavior}
\end{figure}

An example of temporal evolution of the system is shown in FIG \ref{fig_model_detail}. 
If all the species are in active state and have positive fitnesses, nothing will happen. Therefore we call such a state as a {\it persistent state}. 
In the previous models, we added a new species every time the community has reached a persistent state. This corresponds to a low-introduction (mutation, invasion, etc) rate limit. 
In the present model, however, it is also possible that the system relaxes to a limit cycle and never reaches a persistent state (FIG \ref{fig_blinking_behavior}). 
Therefore, we need a new parameter for the time interval of the species introduction, $T_{\rm int}$. 
In the following, we take a long interval: $T_{\rm int} = 100$ to keep a low-introduction rate, unless otherwise noted.

\section{\label{sec:level3}Results}
Following the approach of our previous study, we assess the robustness of the emergent system by the long-term trend of the system size, i.e. the number of species, under the successive introduction of new species. In our original model without any dormant mechanism, the system can grow limitlessly thus it is robust enough against the inclusion of new species, if the number of interactions given for each newly introduced species, $m$, is kept within a moderate range, i.e., $5 \le m \le 18$. In contrast, the system with $m$ outside this range, keeps fluctuating with a finite size. 
These fluctuations may lead to the extinction of the entire system and the lower the mean level is the higher is the probability for such an event. 
To avoid this possibility, we adopt an incubation rule when the system size becomes smaller than the initial system size $N_0$. 
Under the incubation rule, we let totally isolated species (i.e. $f_i = 0$) stay in the active state or inactive state. This treatment prevents the total collapse of the system and provides the system with many more opportunities to search for growth from different initial conditions.

\begin{figure}[bth]
\centering 
\includegraphics[width=0.4\textwidth]{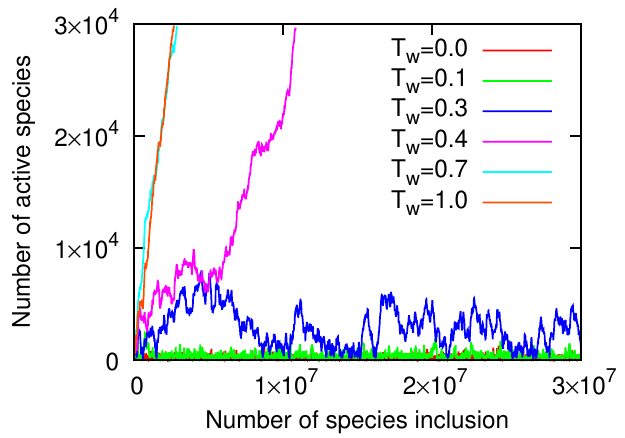}
\caption{
The temporal evolutions of total number of active species $N_{\mbox{active}}(t)$ under the successive introduction of new species with $m = 25$ interactions.
The unit for time is $T_{\rm int}$ i.e. the horizontal axis corresponds to the accumulated number of introduced species.
The size of the emergent system diverges in time if the waiting time of dormancy is long ($T_W \ge 0.4$) while it fluctuates within a finite size for shorter waiting time ($T_W \le 0.3$). 
}
\label{fig_dynamics_n_active}
\end{figure}

For sufficiently large initial system size, typically $N_0 \ge 100$, the limitless growth and finite size fluctuation behaviour are confirmed to be independent of the initial network structure. 
Therefore, we call the former behaviour taking place in the ``diverging phase'' and the latter in the ``finite phase'' of the parameter space. 
The temporal evolution of the system size of the present model with $m = 25$ is shown in FIG \ref{fig_dynamics_n_active}. 
Inheriting the nature of our original model, the system with short dormancy limit $T_W$ is found to be in the finite phase. 
However, as $T_W$ increases (to the value $T_W = 0.3$) the typical system size shows a clear increase yet it stays finite and for $T_W = 0.4$ and above the system has crossed a certain threshold to show diverging behaviour. 
This clearly illustrates that our newly introduced parameter $T_W$, the time limit for the continuous dormancy, can change the robustness of the system.

\begin{figure}[bth]
\centering 
\includegraphics[width=0.5\textwidth]{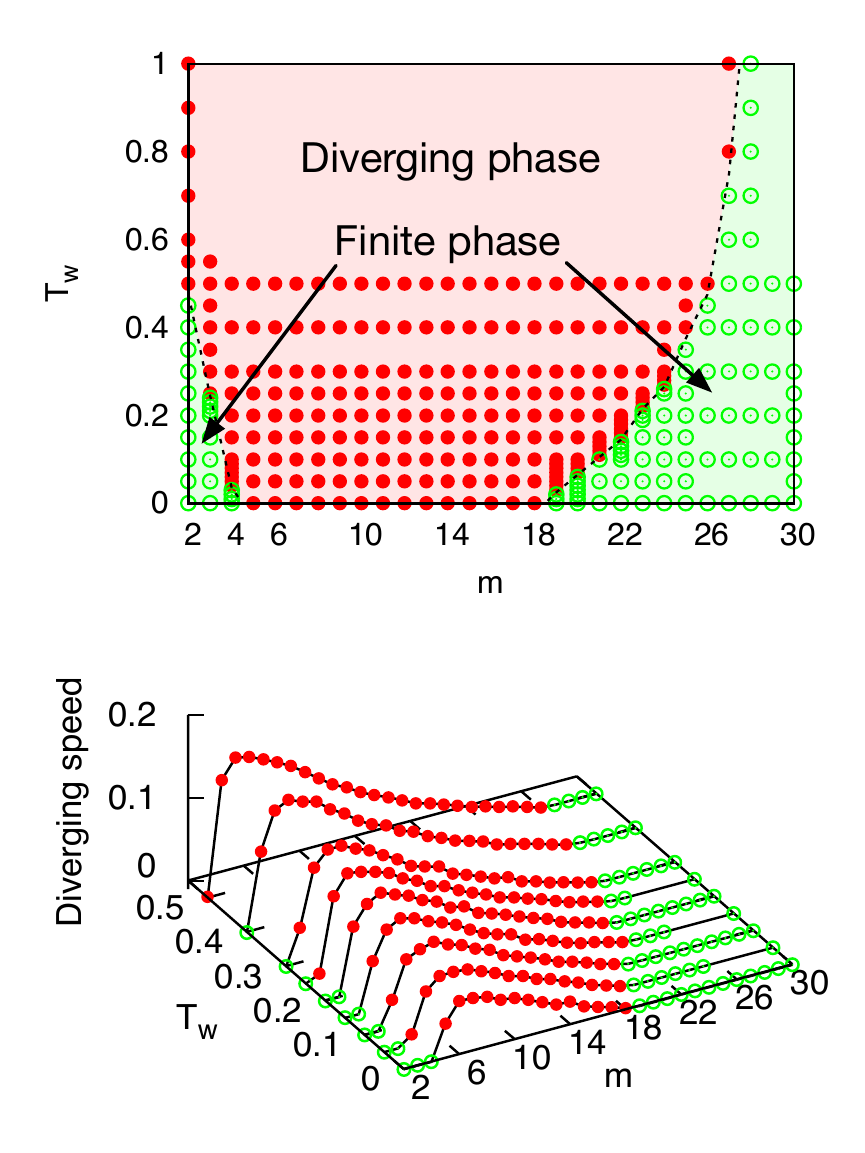}
\caption{
The phase diagram of the evolving open system with dormancy and revival processes. (Top): The speed of divergence
$\displaystyle v = \lim_{t \to \infty} N(t)/t$
for the given original and new key parameters, $m$ and $T_W$.
The points where $v$ is evaluated to be positive are shown by filled red symbols.
(Bottom): The corresponding phase diagram.
}
\label{fig_phase_diagram}
\end{figure}

Next we will explore the whole phase diagram with systematic computer simulations by scanning through the $m$ vs. $T_W$ parameter space. The obtained phase diagram is shown in FIG \ref{fig_phase_diagram}, where it is seen that the introduction of dormancy and revival processes broaden the diverging phase. 
While this effect turns out to be larger for longer dormancy time limit $T_W$, yet it is not possible to get the system with very dense interactions ($m \ge 28$) to the diverging phase.

\begin{figure}[bth]
\centering 
\includegraphics[width=0.4\textwidth]{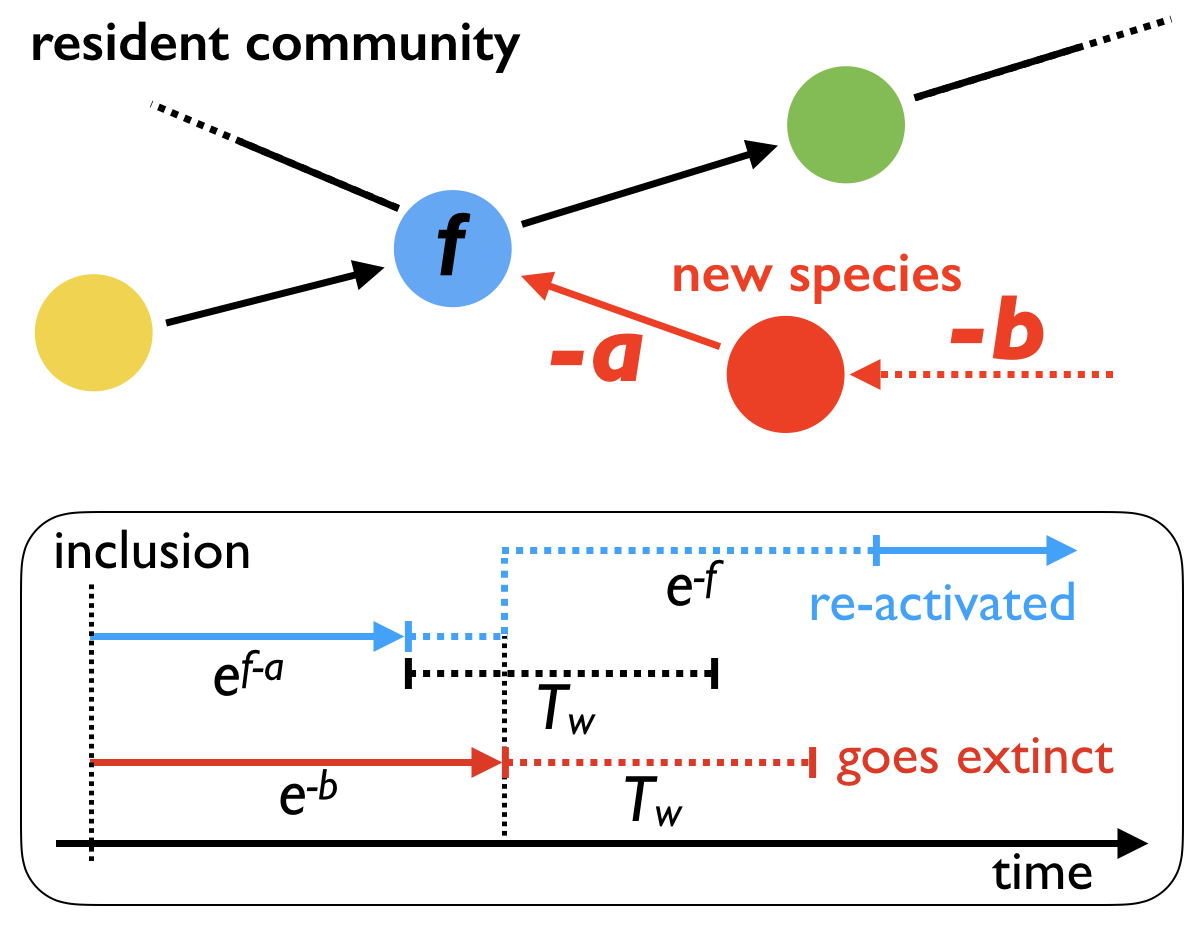}
\caption{
The mechanism of rejecting the attack by species with non-positive fitness.
}
\label{fig_rejection_mechanism}
\end{figure}

The main mechanism of this enforcement is the rejection of ``falling-together-attacks''.
To illustrate this, let us consider a situation that a negative link weight ($-a$) is added to a resident species by a newly introduced species, which has zero or negative fitness value, $-b$ (FIG \ref{fig_rejection_mechanism}).

In our original model, in which the least fit species goes extinct first, the attacked resident species and the new species sequentially go extinct for $f-a < -b$ and otherwise only the new species goes extinct (i.e. is rejected).
Especially for the newly introduced species with no incoming links ($b=0$, solitary attack), every attack strong enough ($f < a$) can kill the resident species before the newly introduced attacker species goes extinct.

In the present model the situation is different as the resident species has another chance to reject such a falling-together attack. The rejection happens if the resident species can survive in the inactivated state until the newly added species stays inactivated. The condition for this type of dynamics is as follows
\begin{equation}
	f-a < -b < \ln \big( {\rm e}^{f-a} + T_W \big).
	\label{eq_rejection_range}
\end{equation}
Therefore, even a strong attack ($f > a$) by a solitary new species ($b=0$) is rejected if $T_W > 1- {\rm e}^{f-a}$. And if $T_W \ge 1$, i. e. the limit of the dormancy period is long enough, even the solitary attacks never become successful. 
Note that the rejection acts perfectly in a special case of $m=1$, because in this situation every inclusion of new species corresponds to either a solitary attack or an attachment of species with no outgoing link. Therefore, even for this most sparse condition, large $T_W$ drives the system with a mutually supporting community core to grow infinitely in size. 
However, such a growth is highly dependent on the initial condition (if there is no core in the initial network, the system collapses) which is out of the scope of this study. 
Thus we excluded this case from the phase diagram.

\begin{figure}[bth]
\centering 
\includegraphics[width=0.5\textwidth]{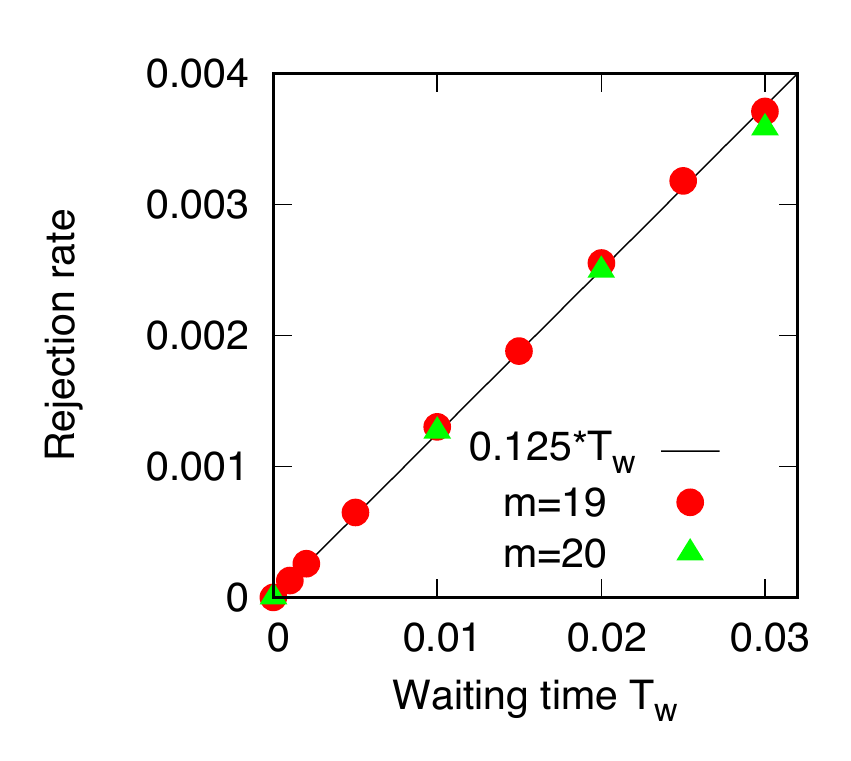}
\caption{
The rejection rate obtained from the simulation in the dense regime. In the small $T_W$ regime shown here, the rejection rate increases linearly to $T_W$.
}
\label{fig_rejection_rate}
\end{figure}

The increment of probability to reject falling-together-attacks directly contributes to the growth rate of the system, $v = N(t)/t$. 
A rough estimate of it near the upper phase boundary ($m \sim 18$) predicts a linear increase of the rejections to $T_W$ for the small $T_W$ regime (see Appendix for details), which is confirmed in the simulation (see FIG. \ref{fig_rejection_rate}). 
The observed contribution of the additional rejections to the system's growth rate, $\Delta v \sim T_W / 8$, predicts the slope of the phase boundary to behave as $\displaystyle \Delta m^* \sim 20 \ T_W$. This is found to be consistent with the phase portrait.

\begin{figure}[bth]
\centering 
\includegraphics[width=0.4\textwidth]{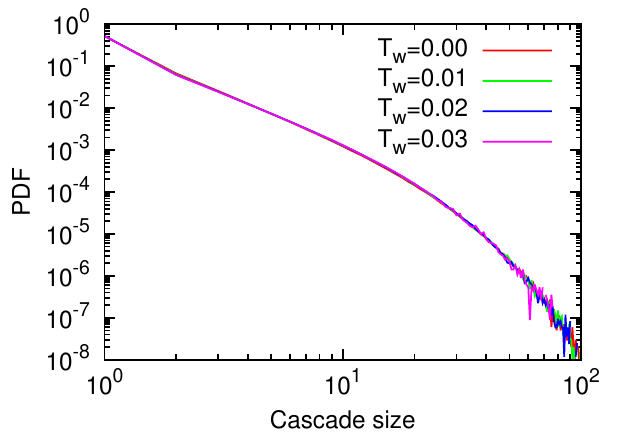}
\caption{
The cascade size distributions of the extinctions in the model with $m=4$.
The distributions from the systems in the finite phase ($T_W \le 0.02$) and from the diverging phase ($T_W = 0.03$) overlap well each other, indicating the structure of the emerging networks is kept.
}
\label{fig_cascadesize}
\end{figure}

The effect of rejections in the sparse regime ($m \le 4$) needs to be estimated differently. 
This is because the probability to have a solitary attack is larger. What is more significant, however, is the fact that the resident community has a sparse network structure, which in turn is very prone to a loss of certain species and can cause a cascade of extinctions of species supported by that species. Therefore, the effect of the increased chance of rejection can be more drastic. It is also possible that the structure of the emergent networks is changed, although the well kept distributions of extinction cascade size suggests it to be negligible at least for $m=4$ (FIG. \ref{fig_cascadesize}). 
The consideration above predicts the broadening of the diverging phase, but it is difficult to give an estimate of the effect of $T_W$ against the very steep drop of the growth rate in this regime of the phase diagram.
\vspace*{-5pt}
\vspace*{-7pt}

\section{\label{sec:level4} Summary and Discussion}
We have studied the robustness of an evolving system against successive inclusions of new elements or constituents, each with an ability to survive temporarily under unfavourable conditions in the state of being inactive. 
It is found that the introduction of the inactivation and revival processes broadens the phase the systems stays robust. 
This reinforcement of the emerging system is mainly due to its increased ability to reject falling-together type attacks. 
It should be noted that the broadening of the robust phase has a limit: systems with $m \ge 28$ stay in the finite phase even at $T_W = 1$, where the rejection probability reaches its maximum. 
The short term rejection process, in which a possible extinction of a species caused by the attack from a species with poor fitness is altered by the extinction of the attacker, can be regarded as a simplified dynamics in a class of population dynamics models~\cite{Taylor1988A, Taylor1988B, Tokita1999, Kondoh2003}. 
Because another type of interaction form, namely the ratio-dependent interaction~\cite{RatioDependentPreying_JTB1989}, is known to reduce to our previous model~\cite{Shimada2015SMSEC}, the extension of the model in this study has broadened the applicability of our theoretical framework. 
Similarly to our earlier results~\cite{Shimada2014SREP, Ogushi2017SREP}, 
we have found that the number of interactions per species limits the system's robustness. There are empirical findings in support to this observation~\cite{SparsityOfFW_Ings2009}.

As for the modelling in general the population dynamics models based on differential or difference state equations are able to describe rich evolutionary patterns following periodic and even chaotic trajectories, as observed in nature~\cite{MicrobialChaos2005, PlanktonChaos2008}. However, this approach is generally computationally so costly that larger system sizes and longer time scales could not be studied. In order to circumvent these problems we have taken a network based approach, which is able to describe the dynamics of the system over much longer evolutionary time scale.

Although our present analysis covers up to the long-dormancy time limit ($T_W = 1$) in terms of the resulting short term rejection process, far longer dormancy limit ($T_W \gg T_{\rm int}$) could bring new phenomena. Under such condition, inactive species can survive evolutionary time scale during which new species are introduced and that change the community. 
In some cases and for various kinds of systems, such as biological, social, and economic systems, it may be important to consider such long dormancy periods~\cite{LongLivingBacteria2002}. 
Also, the effect of bidirectionality~\cite{Ogushi2017SREP} of the interaction should be examined, because it is expected to make the emergent system to show limit cycles more frequently. 
These two regimes, although that require heavier computation power, will reveal new phenomena and will better bridge with the continuous time dynamics models. 
Extending our approach so that some aspects of short term dynamics of more complex models is kept, with further spacial extension focusing on some aspects hardly accessible by traditional methods, is a promising way to treat evolutionary problems better~\cite{ExtinctionDept_Kuussaari2009, ER_Gonzalez2012}.

\section{\label{sec:level6}Appendix}

\subsection{Model procedure}
    \renewcommand{\labelenumi}{(\arabic{enumi})}
    \renewcommand{\labelenumii}{(\roman{enumii})}
    \renewcommand{\labelenumiii}{(\alph{enumiii})}
\begin{enumerate}
	\setcounter{enumi}{-1}
	\item  (Create an initial system)
    \begin{enumerate}
    	\item Prepare $N_{0}$ species and connect them randomly by $L_0$ unidirectional links with link weights denoted by $a_{ij}$. Typical settings are $N_0 = 100$ and $L_0 = 10 N_0$.
    	\item All species have its state variable ($S_i = \{ -1, 1 \}$, $1$ and $-1$ denote active and inactive states, respectively), the time counters for state change $g_i$, and the counter for extinction $h_i$. Those are set to the initial values: $\{S_i\} = 1, \ \{g_i\} = 1, \ \{h_i\} = T_W$.
        \item Set the system time at $t = 0$ and the time for the next new species introduction $T_{\rm next} = T_{\rm int}$. 
    \end{enumerate}
    
	\item Calculate the fitness $f_{i}$ of each species,
		\begin{equation*}
			f_i = \sum_{j}^{incoming} \left( \frac{1+S_j}{2} \right) a_{ij}. 
		\end{equation*}
    \item Reset the time counter if needed:
    \begin{equation*}
    \begin{cases}
        g_i = 1 & \text{($S_i = +1$ and $f_i > 0 \ \cap \ f^{\rm old}_i \le 0$)}\\
        g_i = 1 & \text{($S_i = +1$ and $f_i \le 0 \ \cap \ f^{\rm old}_i > 0$)}\\
        g_i = 1 & \text{($S_i = -1$ and $f_i > 0 \ \cap \ f^{\rm old}_i \le 0$)}\\
        h_i = T_W & \text{($S_i = -1$ and $f_i \le 0 \ \cap \ f^{\rm old}_i > 0$)} 
    \end{cases}
	\end{equation*}
    where $f^{\rm old}_i$ is the fitness at the previous time step.
	\item Calculate the remaining time till the next event for each species, $\delta t_i$: 
		\begin{equation*}
			\delta t_i = 
			\begin{cases}
				T_{\rm next} - t  & (S_i = +1, f_i > 0: \ \mbox{no state change})\\
				g_i \ {\rm e}^{S_i f_i} & (S_i = +1, f_i \le 0: \ \mbox{inactivation})\\
				g_i \ {\rm e}^{S_i f_i} & (S_i = -1, f_i > 0: \ \mbox{reactivation})\\
                T_{\rm next} - t  & (S_i = -1, f_i \le 0: \ \mbox{extinction}).
			\end{cases}
		\end{equation*}
        
    \item Find the shortest time to the next event in the system: \ $\delta t^*_j = \min \{\delta t_i \}$.
    
	\item Time translation of the system from $t$ to $t + \delta t^*_j$ 
    \begin{enumerate}
    	\item Update the system time $t = t + \delta t^*_j$
        \item Update the time counters:
        \begin{equation*}
			\begin{cases}
        		g_i = g_i & \text{($S_i = +1$ and $f_i > 0$)}\\
       			g_i = g_i - {\rm e}^{S_if_i} \delta t^*_j & \text{($S_i=+1$ and $f_i\le 0$)}\\
        		g_i = g_i - {\rm e}^{S_if_i} \delta t^*_j & \text{($S_i=-1$ and $f_i> 0$)}\\
        		h_i = h_i-\delta t^*_j & \text{($S_i = -1$ and $f_i \le 0$)} 
			\end{cases}
		\end{equation*}
        
        \item Extinction: \
        If $h_i \le 0$, delete the species $i$ and all links connecting to and from it.
     \end{enumerate}
     
     \item Treat the event at $t$ (state change of species $j$ or new species introduction)
        \begin{itemize}
            \item If $t<T_{\rm next}$, treat the nearest state change of species, $j$:
            \begin{enumerate}
        		\item Update the state of the species $j$:\\ $S_j = -S_j$.
            \item Reset the time counters:\\ $g_j = 1$ and $h_j = T_W$. 
            \end{enumerate}
            \item If $t=T_{\rm next}$, add a new species:
            \begin{enumerate}
            	\item The new species is added in active state ($S = +1$) with the time counters $g = 1$ and $h = T_W$. 
        \item $m$ interacting species are randomly chosen from the resident species.
        \item The new species forms $m$ directed unidirectional links. The direction of each new link is chosen with a equal probability $1/2$.
        \item The link weights are also randomly chosen from a standard normal distribution.
        \item Update the time for the next species introduction: \quad $T_{\rm next} = T_{\rm next}+T_{\rm int}$.
            \end{enumerate}
        \end{itemize}

    \item Recalculate the fitness: go back to step (1).
\end{enumerate}

\vspace{5mm}
\subsection{Estimation of the rate of the additional rejections and its effect}
Here we first roughly estimate the increment of the chance to reject such falling-together-attack which directly contributes to the growth rate of the system, $v = N(t)/t$, near the upper phase boundary ($m \sim 18$).
In the vicinity of the phase boundary in the dense regime, an inclusion of new species causes one strong attack ($f < a$) event in average. The distribution of $f-a$ is given by the negative side of the convolution:
\begin{equation}
	\rho(f-a) = \int_0^\infty \bar{f}(\xi) \ G(1, f-a - \xi ) \ d\xi,
\end{equation}
where $\bar f(x)$ and $G(\sigma, x)$ represent the equilibrium fitness distribution of the emergent system and the Gaussian distribution with its standard deviation $\sigma$, respectively.
The distribution of the fitness of newly added species, $-b$,  is well approximated by the negative half side of the Gaussian distribution $G(\sqrt{m/2}, -b)$, where $m/2$ is the average number of incoming links.
For small $T_W$,
the condition to have the dormancy-aided rejection, Eq.(\ref{eq_rejection_range}), is 
\begin{equation}
	f-a < -b < f-a + \frac{T_W}{{\rm e}^{f-a}}.
	\label{eq_rejection_range_linear}
\end{equation}
Substituting $\rho(-b)$ near $0$ by its peak value $G(\sqrt{m/2}, 0) = 1/\sqrt{\pi m}$, and taking $f-a \sim - 1/2$ as a typical attack strength, an estimated increment in the system's growth rate brought by the increase of the rejection is
\begin{equation}
	\Delta v_{est.}\sim \sqrt{\frac{\rm e}{\pi m}} T_W \sim \frac{T_W}{5}. 
\end{equation}
We can confirm this linear relation between the rejection rate $\Delta v$ and $T_W$ in the simulation results for $m = 19$ and $20$ (FIG. \ref{fig_rejection_rate}). And the observed slope
\begin{equation}
	\Delta v_{obs.} = \frac{T_W}{8}
\end{equation}
is also consistent with the very rough estimation above. 

Taking the linear slope of the system's intrinsic growth rate to $m$ obtained from the observed growth rates,
\begin{equation}
	\frac{\Delta v}{\Delta m} \sim \frac{0.06}{10},
\end{equation}
we reaches to an estimation for the slope of phase boundary
\begin{equation}
	\Delta m \sim 20 \ T_W.
\end{equation}

\vskip6pt

\enlargethispage{20pt}

\section*{Ethics}
This study did not require ethical approval.
\section*{Author's Contributions}
F.O. and T.S. conceived the model and conducted the simulation. All authors analysed the results and wrote the manuscript.
\section*{Competing Interests}
We declare we have no competing interests. 
\begin{acknowledgments}
F.O. was partly supported by "Materials Research by Information Integration" Initiative (MI2I) project of the Support Program for Starting Up Innovation Hub from the Japan Science and Technology Agency (JST). K.K. acknowledges financial support by the Academy of Finland Research project (COSDYN) No. 276439, EU HORIZON 2020 FET Open RIA project (IBSEN) No. 662725, EU HORIZON 2020 INFRAIA-1-2014-2015 program project (SoBigData) No. 654024, and the Rutherford Foundation Visiting Fellowship at The Alan Turing Institute, UK. JK thanks for hospitality of Aalto University.
T.S. was partly supported by JSPS KAKENHI Grant Number 15K05202 and 18K03449.\end{acknowledgments}
\section*{Disclaimer}
Any opinions, findings or conclusions are those of authors.

\end{document}